\documentclass[11pt,epsf]{article}
\usepackage{lmodern}
\usepackage[T1]{fontenc}
\usepackage[latin9]{inputenc}
\usepackage{geometry}
\usepackage{amsmath}
\usepackage{amssymb}

\usepackage{graphicx}

\newtheorem{theorem}{Theorem}

\newcommand {\dfn} {\stackrel{\Delta} {=}}
\newcommand {\exe} {\stackrel{\cdot} {=}}

\newcommand{\ttt}{\tilde{\theta}}
\newcommand{\tsig}{\tilde{\sigma}}

\newcommand{\ct}{\hat{\theta}}
\newcommand {\reals} {{\rm I\!R}}

\newcommand {\bx} {\mbox{\boldmath $x$}}
\newcommand {\by} {\mbox{\boldmath $y$}}

\newcommand {\bE} {\mbox{\boldmath $E$}}

\newcommand {\bX} {\mbox{\boldmath $X$}}
\newcommand {\bY} {\mbox{\boldmath $Y$}}

\newcommand{\calA}{{\cal A}}

\newcommand{\calE}{{\cal E}}

\newcommand{\calN}{{\cal N}}

\newcommand{\calQ}{{\cal Q}}

\newcommand{\calX}{{\cal X}}

\begin{document}
\thispagestyle{empty}
\title{Lower Bounds on Exponential Moments of the Quadratic Error in Parameter
Estimation}
\author{Neri Merhav
}
\date{}
\maketitle

\begin{center}
The Andrew \& Erna Viterbi Faculty of Electrical Engineering\\
Technion - Israel Institute of Technology \\
Technion City, Haifa 32000, ISRAEL \\
E--mail: {\tt merhav@ee.technion.ac.il}\\
\end{center}
\vspace{1.5\baselineskip}
\setlength{\baselineskip}{1.5\baselineskip}

\begin{center}
{\bf Abstract}
\end{center}
\setlength{\baselineskip}{0.5\baselineskip}
Considering the problem of risk--sensitive parameter estimation, we propose
a fairly wide family of lower bounds on the exponential moments of the quadratic error, both
in the Bayesian and the non--Bayesian regime. This family of bounds, which is based on a change
of measures, offers considerable freedom in the choice of the reference
measure, and our efforts are devoted to explore this freedom to a certain
extent. Our focus is mostly on signal
models that are relevant to communication problems, namely, 
models of a parameter--dependent signal 
(modulated signal) corrupted by additive white Gaussian noise, but
the methodology proposed is also applicable to other types of parametric families,
such as models of linear systems driven by random input signals (white noise,
in most cases), and others.
In addition to the well known motivations of the risk--sensitive cost
function (i.e., the exponential quadratic cost function), 
which is most notably, the robustness to
model uncertainty, we also view this cost function as a tool for studying
fundamental limits concerning the tail behavior of the estimation error. 
Another interesting aspect, that we demonstrate
in a certain parametric model, is that the risk--sensitive
cost function may be subjected to phase transitions, 
owing to some analogies with statistical mechanics.\\

\vspace{0.2cm}


\setlength{\baselineskip}{1.5\baselineskip}
\newpage

\section{Introduction}

While the minimum mean square error (MMSE) has been
the most common figure of merit for measuring the performance 
of estimators in several areas, such as 
estimation theory, information/communication theory,
and statistics, by contrast, the exponential moments of the quadratic error,
namely, $\bE\exp\{\alpha(\ct-\theta)^2)\}$ ($\alpha > 0$ being a prescribed
constant, $\theta$ -- a parameter, and $\ct$ -- its estimator), have received much less
attention in those fields.
In the realm of the theory of
optimization and stochastic filtering 
and control, on the other hand, the problem of
minimizing exponential
moments of the quadratic error, 
or exponential moments of any other loss function, 
has been studied rather intensively, and it is well--known as the {\it
risk--sensitive} or {\it risk--averse} cost function (see, e.g.,
\cite{DPMR96}, \cite{DMS99},
\cite{FHH97}, \cite{HM72}, \cite{Whittle81}, \cite{Whittle90} and many
references therein). 

One of the main motivations for using the
exponential function of the loss is to
impose a penalty, or a risk, that is extremely sensitive to large
values of the loss, hence the qualifier ``risk--sensitive'' in the name
of this criterion. 
Another motivation is associated with robustness properties of the
resulting risk--sensitive solutions to model uncertainties \cite{ACD15}, \cite{DJP00}.
Indeed, as explained in \cite[p.\ 349, the paragraph of eq.\
(11)]{p143}, minimization of the exponential moment of the loss function is
equivalent to minimization of the expected loss for the worst distribution in
the $\epsilon$--neighborhood (in the Kullback--Leibler divergence sense) 
of the given (nominal) probability distribution, where $\epsilon$ grows with
the risk--sensitive factor parameter $\alpha$.
There are, in fact, a few additional motivations for
minimizing exponential moments, which are also
relevant to the problem of parameter estimation considered here, and indeed,
risk--sensitive parameter estimation has been studied to a considerable
extent (see, e.g., \cite{BS98}, \cite{Ford99}, \cite{HP13}, \cite{JBE94}, \cite{MED97},
\cite{RSS14}, \cite{RM02}, \cite{YB09},
and references therein).
First and foremost,
the exponential moment, $\bE\exp\{\alpha(\ct-\theta)^2\}$, as a function of
$\alpha$, is obviously the
moment--generating function of $(\ct-\theta)^2$, and as such, it provides the full
information about the entire distribution of this random variable, not just
its first
order moment. 
Thus, in particular, if we are fortunate enough to find
an estimator that
uniformly minimizes $\bE
\exp\{\alpha(\ct-\theta)^2\}$ for all $\alpha \ge 0$ (and there are examples
that this may be the case), then this is much stronger
than just minimizing the
first moment.
Secondly, exponential moments are intimately related to
large--deviations rate functions (owing to the exponential tightness of the
Chernoff bound), and so, the minimization of exponential
moments may give us an edge on minimizing probabilities of (undesired) large
deviations events of the form $\mbox{Pr}\{|\ct-\theta| \ge \delta\}$ (for some threshold
$\delta$ which is related to $\alpha$), or more precisely,
on maximizing the exponential rate of decay of these probabilities. 

Moreover, an important feature of the estimation error, $\epsilon\equiv\ct-\theta$,
is its tail behavior: how fast can the probability density function (pdf) 
of $\epsilon$ possibly decay as
$|\epsilon|\to\infty$ (assuming the support of $\epsilon$ to be unbounded)?
This is not accomplished, in general, by inspecting a figure of merit like
$\mbox{Pr}\{|\ct-\theta| \ge \delta\}$, for a given $\delta$, 
since the optimal estimator under this
criterion may depend on $\delta$, whereas we might be interested to 
assess the decay rate of the entire distribution tail of a single estimator $\ct$.
It turns out that when the support of $\theta$ is unbounded (and so is
that of $\ct$), then there is normally a critical value of $\alpha$, denoted
$\alpha_{\mbox{\tiny c}}$, such that for every $\alpha < \alpha_{\mbox{\tiny
c}}$, the exponential moment $\bE\exp\{\alpha(\ct-\theta)^2\}$ is finite at
least for some estimators,
whereas, for 
$\alpha\ge \alpha_{\mbox{\tiny
c}}$, it must diverge
for any estimator $\ct$. When this is the case, 
this means that there exists no estimator whose pdf
tail decays faster than a Gaussian tail of the form $\exp(-\alpha_{\mbox{\tiny
c}}\epsilon^2)$. Thus, deriving bounds on $\alpha_{\mbox{\tiny c}}$ is an
important aspect of the risk--sensitive cost function, as we shall see in the
sequel. Interestingly, in a large variety of models that we examine, our
bounds on $\alpha_{\mbox{\tiny c}}$ will turn out to be tight in spite of the
fact that the corresponding bounds to the mean exponential quadratic error
are not always tight.

Our main focus in this paper is in lower bounds to the mean exponential
quadratic error as a function of the risk--sensitive parameter $\alpha$.
We propose a rather wide family of such lower bounds, both
in the Bayesian and non--Bayesian regimes, though most of our emphasis is on
the Bayesian regime. This family of bounds is
based on a change
of measures, combined with the plethora of bounds for the ordinary 
mean square error (MSE), and as such, it offers a considerably large 
freedom in the choice of certain ingredients of these bounds. 
Our efforts are devoted to explore this freedom at least to a certain extent.
The bounds are applied mostly to signal
models that are relevant to communication problems, namely,
models of a parameter--dependent signal
(i.e., a modulated signal) corrupted by additive white Gaussian noise, but
the methodology proposed is also applicable to other types of parametric
families, such as models of linear systems driven by random input signals (white noise,
in most cases), memoryless sources parametrized by their letter
probabilities, Markov sources, parametrized by their state transition probabilities, and so on.
Another interesting aspect, that we demonstrate
(in the Bayesian regime) for 
a very simple parametric model, is that the risk--sensitive
cost function may be subjected to phase transitions,
owing to some intimate analogies with statistical mechanics.

The remaining part of this paper is organized as follows.
In Section 2, we establish notation conventions, define the problem, and
provide a few preliminaries.
In Section 3, we provide our generic lower bounds to the risk--sensitive cost
function for both the Bayesian and the non--Bayesian regimes. These two generic
bounds will serve as the basis for several families of bounds that will be
further developed for various classes of parametric models in Section 4 (the
Bayesian regime) and in Section 5 (the non--Bayesian regime). Also, in the
last subsection of
Section 4, we demonstrate that the 
risk--sensitive cost function may exhibit phase transitions
even in some simple parametric models, like that of the binary memoryless
(Bernoulli) source.

\section{Notation, Definitions, Problem Setting and Preliminaries}

We begin by establishing some notation conventions.
We consider a parametric family of probability functions,\footnote{Probability
density functions (pdfs) in the
continuous case, or probability mass functions (pmfs) in the discrete case.}
$\{P(\by|\theta),~\theta\in\calA\}$, where $\theta$ is parameter taking
values in a parameter set $\calA$, and $\by$ is a set of observations.
For the sake of simplicity of the presentation, throughout most of
of this paper, $\theta$ will be a scalar parameter, and accordingly, 
$\calA$ will be either an interval, or half of the real line, or the entire real line.
In a few places along the paper, however, we will let $\theta$ be a vector of finite dimension
$k$ and accordingly, $\calA$ will be $\reals^k$ or a subset of it. The
set of observations, $\by$, may either be a vector of dimension $n$,
$\by=(y_1,\ldots,y_n)$ (in the discrete--time case) or a waveform 
$\by=\{y(t),~0\le t\le T\}$ (in the continuous--time, Gaussian case), depending
on the context. In the latter case, the conditional pdf $P(\by|\theta)$ will be understood
to be defined via a complete family of orthonormal basis functions, according
to the well known conventions (see, e.g., \cite[Chap.\ 8]{Gallager68}). 
In the Bayesian setting, we will assume that
$\theta$ is a random variable, distributed according to a given prior
$P(\theta)$, whose support is $\calA$. The joint density of $\theta$ and $\by$
will then be given by $P(\theta,\by)=P(\theta)P(\by|\theta)$. Alternative
joint densities of $\theta$ and $\by$ will be denoted by the letter $Q$ (e.g.,
$Q(\theta)$, $Q(\by|\theta)$, etc.). 

An estimator, $\ct$, of the parameter
$\theta$, is any measurable function of
$\by$ only. Following customary notation conventions, we will use capital
letters to designate randomness. Accordingly, $\bY$ will denote the random
observation set, and in the Bayesian setting $\Theta$ will denote a random
parameter governed by the prior $P(\theta)$. 

In the non--Bayesian setting,
where $\theta$ is assumed an unknown deterministic variable,
$P_\theta\{\calE\}$ will denote the probability of an event
$\calE$, associated with $\bY$, under $P(\cdot|\theta)$, and similarly, $\bE_\theta\{f(\bY)\}$
will denote the expectation of some function $f$ of the random observation set
$\bY$, with respect to (w.r.t.) $P(\cdot|\theta)$. An estimator $\ct\equiv
\ct(\by)$ is
called {\it unbiased} if for every $\theta\in\calA$, we have $\bE_\theta\ct=\theta$.
The Kullback--Leibler divergence between the conditional densities
$P_{\ttt}=P(\cdot|\ttt)$ and $P_{\theta}=P(\cdot|\theta)$ will be defined as
\begin{equation}
D(P_{\ttt}\|P_\theta)=\bE_{\ttt}\ln\frac{P(\bY|\ttt)}{P(\bY|\theta)},
\end{equation}
provided that the support of $P(\cdot|\ttt)$ covers the one of
$P(\cdot|\theta)$.

In the Bayesian setting, the probability of an event $\calE$, associated with
$\bY$ and $\Theta$, will be denoted by
$P\{\calE\}$ and the expectation of a given function $f(\bY,\Theta)$, will be
denoted by $\bE\{f(\bY,\Theta)\}$. The conditional expectation of
$f(\bY,\Theta)$ given $\bY=\by$ (which is the same as the conditional
expectation of $f(\by,\Theta)$) will be denoted by $\bE\{f(\by,\Theta)|\by\}$.
Expectations and conditional expectations w.r.t.\ an alternative
joint density $\{Q(\theta,\by)\}$ will be subscripted by $Q$, i.e.,
$\bE_Q\{f(\Theta,\bY)\}$, $\bE_Q\{f(\Theta,\by)|\by\}$, etc.
The Kullback--Leibler divergence between
$Q$ and $P$ will be defined as
\begin{equation}
D(Q\|P)=\bE_Q\ln\frac{Q(\Theta,\bY)}{P(\Theta,\bY)},
\end{equation}
provided that the support of $Q$ covers the support of $P$.

In this paper, we judge the performance of 
any estimator according to the exponential moment of the squared error, henceforth
referred to as the {\it risk--sensitive cost function}, which is parametrized
by a positive real $\alpha$, called the {\it risk--sensitive factor}, and is
defined as follows.
In the non--Bayesian
regime, 
\begin{equation}
\Lambda_{\mbox{\tiny NB}}
(\ct,\theta,\alpha)\dfn \ln \bE_\theta\left\{\exp\left(\alpha
[\ct(\bY)-\theta]^2\right)\right\},
\end{equation}
with the quest of minimizing $\Lambda_{\mbox{\tiny NB}}$ over all unbiased
estimators, uniformly for all $\theta\in\calA$.
In the Bayesian regime, the risk--sensitive cost function is defined as
\begin{equation}
\Lambda_{\mbox{\tiny B}}(\ct,\alpha)\dfn \ln \bE\left\{\exp\left(\alpha
[\ct(\bY)-\Theta]^2\right)\right\},
\end{equation}
with the quest of minimizing $\Lambda_{\mbox{\tiny B}}$ over all
estimators in general. 
The risk--sensitive factor, $\alpha$, controls the degree of
risk--sensitivity, or the degree of robustness of the estimator that minimizes
the risk--sensitive cost function. The larger is $\alpha$, the greater is the
sensitivity to large errors. The limit $\alpha\to 0$ recovers the case of ordinary 
MSE estimation.

It should be pointed out that even in the Bayesian regime (let alone the
non--Bayesian one), the optimal estimator 
\begin{equation}
\ct=\mbox{arg}\min_\eta
\bE\{e^{\alpha(\Theta-\eta)^2}|\by\}
\end{equation}
is not trivial to calculate, in general, as it is
associated with the solution $\eta$ of the equation
\begin{equation}
\eta=\frac{\bE\{\Theta e^{\alpha(\Theta-\eta)^2}|\by\}}
{\bE\{e^{\alpha(\Theta-\eta)^2}|\by\}},
\end{equation}
which cannot be solved in closed--form in most cases.
This is even more difficult than calculating the MMSE estimator, i.e., the
conditional expectation,
$\bE\{\Theta|\by\}$, which is known to be a non--trivial task (if not
completely undoable) on its own
right, in the vast majority of cases of practical interest.
Thus, the need for good lower bounds to $\Lambda_{\mbox{\tiny
NB}}(\ct,\theta,\alpha)$ and
$\Lambda_{\mbox{\tiny B}}(\ct,\alpha)$ is at least as crucial as in the
classical case of the MSE cost function.

\section{Generic Lower Bounds}

Our basic, generic lower bounds, for both the Bayesian and the 
non--Bayesian regimes, are provided in the following theorem, whose very
simple proof appears at the end of this section.

\begin{theorem}
Consider the parametric family defined in Section 2.
\begin{enumerate}
\item For every estimator $\ct$,
\begin{equation}
\label{genericb}
\Lambda_{\mbox{\tiny B}}(\ct,\alpha)\ge \alpha L_{\mbox{\tiny B}}(Q)-D(Q\|P),
\end{equation}
where $Q$ is an arbitrary joint density of $(\Theta,\bY)$ and
$L_{\mbox{\tiny B}}(Q)$ is an arbitrary lower bound to $\bE_Q[\ct(\bY)-\Theta]^2$.
\item For every unbiased estimator $\ct$,
\begin{equation}
\label{genericnb}
\Lambda_{\mbox{\tiny NB}}(\ct,\theta,\alpha)\ge
\alpha L_{\mbox{\tiny
NB}}(\ttt)+\alpha(\ttt-\theta)^2-D(P_{\ttt}\|P_\theta),
\end{equation}
where $\ttt$ is an arbitrary parameter value in $\calA$ and
$L_{\mbox{\tiny NB}}(\ttt)$ is any non--Bayesian lower bound on the MSE,
$\bE_{\ttt}(\ct-\ttt)^2$, 
for unbiased estimators.
\end{enumerate}
\end{theorem}

As can be seen, both generic bounds offer a rather wide spectrum of
choices with many degrees of freedom. In
the Bayesian lower bound (part 1 of Theorem 1), 
one has the freedom to choose both the joint density $Q$, henceforth
referred to as the {\it reference model}, and the
MSE lower bound,
$L_{\mbox{\tiny B}}(Q)$. As for the former, since the inequality
(\ref{genericb}) applies for any
$Q$, one has the freedom to select the one that maximizes the r.h.s.\ of
(\ref{genericb}) over
a certain class $\calQ$ of reference models, and it would be wise to define the
class $\calQ$ to be wide enough to yield good bounds, on the one hand, and
structured enough to make the problem of maximization over $\calQ$ tractable, on
the other hand. First and foremost, $\calQ$ should be a family of other parametric
models for which it is relatively easy to derive a good lower bound,
$L_{\mbox{\tiny B}}(Q)$ (or even the exact MMSE), as well as an expression (or at least an upper
bound) to $D(Q\|P)$.
Of course, the choice $Q=P$ is always legitimate, but it normally leads to a rather
trivial lower bound -- the same bound that is obtained by a simple application of
the Jensen inequality ($\ln \bE e^Z\ge \bE Z$), which is a relatively weak
bound in most cases. Of course, if $\calQ$ is chosen to include $P$ as a
member, the bound resulting from
maximization over $\calQ$ cannot be worse than this trivial lower bound.
The MSE lower bound, $L_{\mbox{\tiny B}}(Q)$, can chosen, 
for example, to be the Bayesian Cram\'er--Rao bound
\cite{vantrees}, or the Weiss--Weinstein lower bound \cite{weissphd},
or any one of their many variants, see, e.g., \cite{vantreesbell} and
many references therein. 

Similar comments apply
to the non--Bayesian setting (part 2 of Theorem 1). Here the degrees of
freedom are in the
selection of the {\it reference parameter value}, $\ttt$, and in the selection
of the non--Bayesian MSE lower bound, $L_{\mbox{\tiny NB}}(\ttt)$. 
The best choice of $\ttt$ is, of course, the one that maximizes the r.h.s.\ of
(\ref{genericnb}) over $\calA$, but this maximization is not always an easy
task. The MSE lower bound, $L_{\mbox{\tiny NB}}(\ttt)$, can
be taken to be one of many existing non--Bayesian lower bounds for unbiased
estimators, for example,
the non--Bayesian Cram\'er--Rao lower bound (see, e.g.,
\cite{vantrees}), the Bhattacharyya bound \cite{bhattacharyya46}, the
Chapman--Robbins bound
\cite{CR51}, the Fraser--Guttman bound \cite{FG52},
the Barankin bound \cite{Barankin49}, the Keifer bound \cite{Keifer52}, etc.

We note that when the support $\calA$ of the parameter $\theta$ is unbounded, 
the maximization of the lower bound (over $Q\in\calQ$ -- in the Bayesian
case, or over $\ttt\in\calA$ -- in the non--Bayesian case) might yield an
infinite value for large enough $\alpha$, say,
for every $\alpha\ge\alpha_{\mbox{\tiny c}}$, where 
$\alpha_{\mbox{\tiny c}}$ is referred to as the critical value of $\alpha$.
When this is
the case, it means that the pdf of the estimation
error, $\epsilon\dfn\ct-\theta$, decays at a rate slower than
$\exp\{-\alpha_{\mbox{\tiny c}}\epsilon^2\}$ for $|\epsilon|\to\infty$.
The lower bounds of Theorem 1 can therefore yield upper bounds to
$\alpha_{\mbox{\tiny c}}$. If $\alpha_{\mbox{\tiny c}}=0$,
it means that there is no estimator with an estimation error whose tail decays
as fast as any Gaussian. We will encounter situations where this is indeed the
case: interestingly, while for many estimators, the error is nearly Gaussian around the
origin (due to an underlying central limit theorem), the tails may decay at a much
slower rate than any Gaussian tail. In the other extreme,
if $\alpha_{\mbox{\tiny c}}=\infty$, which is
the case where the lower bound is finite no matter how large $\alpha$ may be,
the tail of the error may decay faster than any Gaussian. Of course, when
$\calA$ is a finite interval, this is trivially the case. We shall comment on the
behavior of $\alpha_{\mbox{\tiny c}}$ in the various cases that we
consider.

Some of the resulting bounds (Bayesian and non--Bayesian alike) can
easily be extended to the vector case.
Considering, for example, the risk--sensitive cost function
\begin{equation}
\Lambda_{\mbox{\tiny NB}}(\ct,\theta,\alpha)=\ln\bE_\theta\exp\left\{
[\alpha^T(\ct-\theta)]^2\right\},
\end{equation}
where $\alpha$ is $\theta$ are now both column vectors of dimension $k$ and
the superscript $T$ denotes transposition, the above quantity is readily lower bounded
by
\begin{equation}
\alpha^T[I^{-1}(\ttt)+(\ttt-\theta)(\ttt-\theta)^T]\alpha-D(P_{\ttt}\|P_\theta),
\end{equation}
where $I^{-1}(\ttt)$ is the inverse of the $k\times k$ Fisher information matrix
associated with the parametric family $\{P(\cdot|\theta),~\theta\in\calA\}$.
This gives some fundamental limitations on arbitrary projections of the
estimation error as well as on the behavior of the tails of the estimation
error in various directions in $\reals^k$.

In the remaining part of this paper, we shall make an attempt to explore some of these degrees of
freedom. Most of our efforts will be given to the Bayesian regime, but we
will also devote some attention to the Bayesian regime.
We end this section by providing the proof of Theorem 1.\\

\noindent
{\it Proof of Theorem 1.}
We begin from a very simple well--known inequality, which stands at the basis
of the Laplace principle \cite{DE97} or more generally, the Varadhan integral
lemma \cite[Section 4.3]{DZ93}. 
Let $Z$ be a random variable, governed by a probability distribution $P$,
and let $Q$ be an alternative probability distribution on the same space, such
that $D(Q\|P) < \infty$. Now,
\begin{equation}
\label{laplace}
\ln\bE_P e^Z=\ln \bE_Q \exp\left\{Z+\ln\frac{P(Z)}{Q(Z)}\right\}\ge \bE_Q
Z-D(Q\|P),
\end{equation}
where the second step follows from the Jensen inequality.\footnote{Equality is
achieved if the r.h.s.\ is maximized w.r.t.\ $Q$, namely, if $Q(z)$ is taken to be
proportional to $P(z)e^z$ (provided that $P(z)e^z$ is integrable), 
see, e.g., \cite{p143} and references therein.}

Next apply eq.\ (\ref{laplace}) to both the Bayesian and the non--Bayesian
regime. For the Bayesian regime, set
$Z=\alpha[\ct(\bY)-\Theta]^2$, then further lower bound $\bE_Q[\ct(\bY)-\Theta]^2$
by $L_{\mbox{\tiny B}}(Q)$, which will yield part 1 of Theorem 1.
For the non--Bayesian regime 
(part 2 of Theorem
1), set $Z=[\ct(\bY)-\theta]^2$, 
use $P_\theta$ in the role of $P$, $P_{\ttt}$ -- in the role of $Q$, and
then further lower bound $\bE_{\ttt}(\ct-\theta)^2$ by
\begin{equation}
\bE_{\ttt}(\ct-\theta)^2=\bE_{\ttt}(\ct-\ttt)^2+(\ttt-\theta)^2\ge
L_{\mbox{\tiny NB}}(\ttt)+(\ttt-\theta)^2,
\end{equation}
where in the first step, we have used the assumed unbiasedness 
of $\ct$ and in the second step, we have used the definition of
$L_{\mbox{\tiny NB}}(\ttt)$. This completes the proof of Theorem 1.

\section{The Bayesian Regime}

\subsection{Conditions for Tightness of the Lower Bound}

Before exploring the lower bound of the Bayesian regime in specific parametric
models, we would like first to furnish conditions under which this bound
is tight. Applying Observation 1 of \cite[p.\ 347]{p143} to Bayesian parameter
estimation, we find that an estimator $\ct$ minimizes $\Lambda_{\mbox{\tiny
B}}(\ct,\alpha)$ if it also minimizes $\bE_Q(\ct-\Theta)^2$, where
$Q(\theta|\by)$ is given by
\begin{equation}
\label{obs1}
Q(\theta|\by)=\frac{P(\theta|\by)e^{\alpha[\ct(\by)-\theta]^2}}{Z(\by)},
\end{equation}
and where $Z(\by)$ is a normalization constant, so that $Q(\cdot|\by)$ 
would integrate to unity.
In other words, to minimize $\Lambda_{\mbox{\tiny B}}(\ct,\alpha)$,
the estimator $\ct$ has to be the conditional expectation of $\Theta$ given
$\by$ w.r.t.\ $Q$. Note that this condition has a circular character, since
$Q$ depends on $\ct$, which in turn depends on $Q$. Therefore, this condition
for optimality is more useful as a criterion to check whether a given estimator is
optimal than as a tool for actually finding the optimal estimator.
Further, let us lower bound $\bE_Q(\ct-\Theta)^2$ by 
$L_{\mbox{\tiny B}}(Q)$, which is given by the Bayesian Cram\'er--Rao lower
bound \cite{vantrees} w.r.t.\ $Q$, that is,
\begin{equation}
\label{bcrlb}
L_{\mbox{\tiny B}}(Q)=\frac{1}{\bE_Q W^2(
\Theta,\bY)},
\end{equation}
where
\begin{equation}
W(\theta,\by)\dfn\frac{\partial\ln Q(\theta,\by)}{\partial\theta}.
\end{equation}
As shown in \cite[p.\ 73]{vantrees}, a necessary and sufficient condition
for the tightness of the Bayesian Cram\'er--Rao lower bound is that
$Q(\theta|\by)$ would be a Gaussian pdf
whose variance is
given by the r.h.s.\ of (\ref{bcrlb}), independently\footnote{A tighter bound
can be obtained by averaging (over $\bY$) the conditional Bayesian 
Cram\'er--Rao lower
bound given $\by$,
where the expectation at the denominator of (\ref{bcrlb}) is replaced
by the conditional expectation given $\by$. The necessary and sufficient
condition for the achievability of this bound is Gaussianity of the posterior,
where both the conditional mean and the conditional variance are allowed to
depend on $\by$.} of $\by$,
and whose mean is an arbitrary function of $\by$ (not necessarily a linear
function). When
this is the case, then obviously, this
function of $\by$ becomes the optimal MMSE estimator, $\ct(\by)$, 
that achieves the Bayesian Cram\'er--Rao
lower bound w.r.t.\ $Q$.
Combining this condition with (\ref{obs1}), we find that
eq.\ (\ref{genericb}), with $L_{\mbox{\tiny B}}(Q)$ given by (\ref{bcrlb}),
is a tight lower bound when $P(\theta|\by)$ is also
Gaussian with mean $\ct(\by)$ and variance that is independent of $\by$.

\subsection{Non--linear Signal Models and Linear Reference Models}

A very well known special case, where the above conditions concerning $Q$ are satisfied, is the case
where $\Theta$ is a Gaussian random variable, and given $\Theta=\theta$, the
observation set $\bY$ is
Gaussian with a mean given by a linear function of $\theta$ (and then, $\ct(\by)$
is a linear function of $\by$). This motivates us to choose the reference model
$Q$ as a jointly Gaussian density of $\Theta$ and $\bY$. 

Let us then begin from the case where both $P$ and $Q$ are such Gaussian linear models.
In particular, under $P$, let $\Theta\sim \calN(0,\sigma^2)$ and for a given
$\Theta=\theta$, 
\begin{equation}
\label{ref}
y(t)=\theta s(t)+n(t),~~~~~0\le t\le T
\end{equation}
where $T$ is the observation interval,
$n(t)$ is additive white Gaussian noise (independent of $\Theta$)
with two--sided spectral density $N_0/2$, and
$\{s(t),~0\le t\le T\}$ is a given waveform with energy $E_s$.
Denoting 
\begin{equation}
z=\int_0^Ts(t)y(t)\mbox{d}t,
\end{equation}
it is easily seen that
\begin{equation}
P(\theta|\by) = P(\theta|z) \propto
\exp\left\{-\frac{N_0+2\sigma^2E_s}{2\sigma^2N_0}\left(\theta-
\frac{\sigma^2}{\sigma^2E_s+N_0/2}\cdot z\right)^2\right\}.
\end{equation}
Thus, if we define 
\begin{eqnarray}
Q(\theta,\by)&\propto&P(\by)P(\theta|\by)\exp\left\{\alpha\left(\theta-
\frac{\sigma^2}{\sigma^2E_s+N_0/2}\cdot z\right)^2\right\}\\
&\propto&P(\by)\exp\left\{-\left(\frac{N_0+2\sigma^2E_s}{2\sigma^2N_0}-\alpha\right)
\left(\theta-\frac{\sigma^2}{\sigma^2E_s+N_0/2}\cdot z\right)^2\right\}
\end{eqnarray}
then the conditions of \cite[Observation 1]{p143} are satisfied for the conditional mean
estimator,
\begin{equation}
\label{condmean}
\ct= \frac{\sigma^2}{\sigma^2E_s+N_0/2}\cdot z,
\end{equation}
and so, the same estimator minimizes also $\Lambda_{\mbox{\tiny
B}}(\ct,\alpha)$ for every
\begin{equation}
0< \alpha < \alpha_{\mbox{\tiny c}}\dfn\frac{N_0+2\sigma^2E_s}{2\sigma^2N_0}
\equiv\frac{1}{2\sigma^2}+\frac{E_s}{N_0}.
\end{equation}
The resulting value of the minimum achievable 
$\Lambda_{\mbox{\tiny B}}(\ct,\alpha)$ is given by
\begin{equation}
\min_{\ct(\cdot)}\Lambda_{\mbox{\tiny B}}(\ct,\alpha)=
\frac{1}{2}\ln\frac{1}{1-\alpha/\alpha_{\mbox{\tiny c}}},
\end{equation}
and so, indeed the estimator error $\epsilon=\ct(\bY)-\Theta$ has a Gaussian
tail at rate $\alpha_{\mbox{\tiny c}}$ as defined above.

Next, suppose that under $P$, $\Theta\sim \calN(0,\sigma^2)$ and
\begin{equation}
y(t)=x(t,\theta)+n(t),~~~~~0\le t\le T
\end{equation}
where $x(t,\theta)$ is continuous--time waveform (depending on $\theta$),
whose energy is $E_x$ independently\footnote{This assumption concerning the
energy is (at least
nearly) satisfied for many parametric signal models, e.g., when $\theta$ designates
delay, frequency, or phase, etc. In these cases, the Gaussian prior on
$\Theta$ makes sense (as an approximation) as long as its standard deviation,
$\sigma$, is significantly smaller than the size of the interval $\calA$
of the support of the parameter.}
of $\theta$.
Now, under $Q$, let $\Theta\sim N(0,\tsig^2)$ and $y(t)$ be as in (\ref{ref}),
so for the given model $P$, 
we have the freedom to choose $\tsig^2$ and the auxiliary signal $s(t)$.
To apply part 1 of Theorem 1, we have
to derive the expression for $D(Q\|P)$, which decomposes to the sum of two
terms. The first is the divergence between the two priors of $\Theta$, which
is given by
\begin{equation}
D[\calN(0,\tsig^2)\|\calN(0,\sigma^2)]=\frac{1}{2}\left[
\frac{\tsig^2}{\sigma^2}-
\ln\frac{\tsig^2}{\sigma^2}-1\right],
\end{equation}
and the second term is the expected 
divergence between the two Gaussian pdfs of $\by$ given $\theta$.
In general, as can easily be verified, 
the divergence between the probability measures of two noisy signals,
$y(t)=x_1(t)+n(t)$, and
$y(t)=x_2(t)+n(t)$, $0\le t\le T$, with the same noise spectral density
$N_0/2$, is given by $\int_0^T[x_1(t)-x_2(t)]^2\mbox{d}t/N_0$.
Therefore, applying (\ref{genericb}), we have
\begin{eqnarray}
\label{1stbound}
\Lambda_{\mbox{\tiny B}}(\ct,\alpha)&\ge&\alpha
\bE_Q(\ct-\Theta)^2-\frac{1}{2}\left[\frac{\tsig^2}{\sigma^2}-
\ln\frac{\tsig^2}{\sigma^2}-1\right]-\nonumber\\
& &\frac{1}{N_0}\bE_Q[\Theta^2E_s+E_x-2\Theta C_{xs}(\Theta)]\\
&\ge&\frac{\alpha\tsig^2N_0}{N_0+2\tsig^2E_s}-\frac{1}{2}\left[\frac{\tsig^2}{\sigma^2}-
\ln\frac{\tsig^2}{\sigma^2}-1\right]-\nonumber\\
& &\frac{\tsig^2E_s+E_x-2\bE_Q\{\Theta
C_{xs}(\Theta)\}}{N_0},
\end{eqnarray}
where
$$C_{xs}(\theta)\dfn\int_0^Ts(t)x(t,\theta)\mbox{d}t.$$
For a given $E_s$, the best choice of the signal $\{s(t),\le t\le T\}$ 
(in the sense of maximizing the lower obund) is the
one that maximizes 
\begin{equation}
\bE_Q\{\Theta C_{xs}(\Theta)\}=\int_0^Ts(t)\bE_Q\{\Theta x(t,\Theta)\}\mbox{d}t,
\end{equation}
namely,
\begin{equation}
s^*(t)=\frac{\sqrt{E_s}}{\sqrt{\int_0^T[\bE_Q\{\Theta
x(t,\Theta)\}]^2\mbox{d}t}}\cdot\bE_Q[\Theta x(t,\Theta)].
\end{equation}
which when substituted back into (\ref{1stbound}), yields
\begin{eqnarray}
\label{2nd}
\Lambda_{\mbox{\tiny B}}(\ct,\alpha)&\ge&
\frac{\alpha\tsig^2N_0}{N_0+2\tsig^2E_s}-\frac{1}{2}\left[\frac{\tsig^2}{\sigma^2}-
\ln\frac{\tsig^2}{\sigma^2}-1\right]-\nonumber\\
& &\frac{1}{N_0}\left[\tsig^2E_s+E_x-2\sqrt{E_s\cdot\int_0^T[\bE_Q\{\Theta
x(t,\Theta)\}]^2\mbox{d}t}\right]\\
&=&\frac{\alpha\tsig^2}{1+2\lambda\tsig^2}-\lambda\tsig^2-
\frac{1}{2}\left[\frac{\tsig^2}{\sigma^2}-
\ln\frac{\tsig^2}{\sigma^2}-1\right]-\nonumber\\
& &\frac{1}{N_0}\left[E_x-2\sqrt{\lambda N_0\cdot\int_0^T[\bE_Q\{\Theta
x(t,\Theta)\}]^2\mbox{d}t}\right],
\end{eqnarray}
where $\lambda\dfn E_s/N_0$
and the remaining degrees of freedom for maximizing the bound are the
parameters $\lambda$ and $\tsig^2$. Consider first the maximization w.r.t.\
$\lambda$. As a function of $\lambda$, the above lower bound has the form
$$f(\lambda)=\frac{a}{1+b\lambda}-c\lambda+d\sqrt{\lambda},$$
where
\begin{eqnarray}
a&=&\alpha\tsig^2\\
b&=&2\tsig^2\\
c&=&\tsig^2\\
d&=&2\sqrt{\frac{1}{N_0}\int_0^T[\bE_Q\{\Theta x(t,\Theta)\}]^2\mbox{d}t}.
\end{eqnarray}
If $a$ is small and/or $b$, $c$ and $d$ are large (which is the case when $\alpha$
is small and/or $\tsig^2$ is large and/or $N_0$ is small), then the first term
is not important and a good choice of $\lambda$ is the one that maximizes
$-c\lambda+d\sqrt{\lambda}$, namely,
$$\lambda=\frac{d^2}{4c^2}=\frac{\int_0^T[\bE_Q\{\Theta
x(t,\Theta)\}]^2\mbox{d}t}{N_0\tsig^4}.$$
On substituting this into (\ref{2nd}), we obtain
\begin{eqnarray}
\Lambda_{\mbox{\tiny B}}(\ct,\alpha)
&\ge&
\frac{\alpha\tsig^2}{1+\frac{2}{N_0\tsig^2}
\int_0^T[\bE_Q\{\Theta x(t,\Theta)\}]^2\mbox{d}t}
+\frac{\int_0^T[\bE_Q\{\Theta
x(t,\Theta)\}]^2\mbox{d}t}{N_0\tsig^2}-\nonumber\\
& &\frac{1}{2}\left[\frac{\tsig^2}{\sigma^2}-
\ln\frac{\tsig^2}{\sigma^2}-1\right]-\frac{E_x}{N_0}.
\end{eqnarray}
To obtain more explicit results, we now particularize the signal model to
phase modulation:
\begin{equation}
x(t,\theta)=\sqrt{\frac{2E_x}{T}}\cdot\cos(\omega
t+\theta),
\end{equation}
where we have the following:
\begin{eqnarray}
\bE_Q\{\Theta
x(t,\Theta)\}&=&\sqrt{\frac{2E_x}{T}}\cdot\mbox{Re}\left\{e^{j\omega t}\bE_Q[\Theta\cdot
e^{j\Theta}]\right\}\nonumber\\
&=&-\sqrt{\frac{2E_x}{T}}\cdot\mbox{Re}\left\{je^{j\omega
t}\frac{\partial}{\partial s}\bE_Q[
e^{js\Theta}]\bigg|_{s=1}\right\}\nonumber\\
&=&-\sqrt{\frac{2E_x}{T}}\cdot\mbox{Re}\left\{je^{j\omega
t}\frac{\partial}{\partial s}
e^{-s^2\tsig^2/2}]\bigg|_{s=1}\right\}\nonumber\\
&=&\sqrt{\frac{2E_x}{T}}\cdot\tsig^2e^{-\tsig^2/2}\cdot\mbox{Re}\{j e^{j\omega
t}\}\nonumber\\
&=& -\sqrt{\frac{2E_x}{T}}\cdot\tsig^2e^{-\tsig^2/2}\sin(\omega t).
\end{eqnarray}
which yields
\begin{equation}
\int_0^T[\bE_Q\{\Theta x(t,\Theta)\}]^2\mbox{d}t= E_x\tsig^4 e^{-\tsig^2}.
\end{equation}
and so, the lower bound becomes, in this case,
\begin{eqnarray}
\label{phase-est}
\Lambda_{\mbox{\tiny B}}(\ct,\alpha)
&\ge&\frac{\alpha\tsig^2}{1+2E_x\tsig^2
e^{-\tsig^2}/N_0}-\frac{E_x}{N_0}(1-\tsig^2 e^{-\tsig^2})-\nonumber\\
& &\frac{1}{2}\left[\frac{\tsig^2}{\sigma^2}-
\ln\frac{\tsig^2}{\sigma^2}-1\right].
\end{eqnarray}
For example, if $\tsig^2=\sigma^2$, this becomes
\begin{equation}
\Lambda_{\mbox{\tiny B}}(\ct,\alpha)\ge
\frac{\alpha\sigma^2}{1+2E_x\sigma^2
e^{-\sigma^2}/N_0}-\frac{E_x}{N_0}(1-\sigma^2 e^{-\sigma^2}).
\end{equation}
Let us compare this to the lower bound obtained by a simple
application of the Jensen inequality, combined with the Bayesian Cram\'er--Rao
lower bound. The result is
\begin{equation}
\Lambda_{\mbox{\tiny B}}(\ct,\alpha)\ge
\frac{\alpha\sigma^2}{1+\sigma^2E_x/2N_0}.
\end{equation}
When $\sigma^2$ is very large, this yields approximately $2\alpha N_0/E_x$,
whereas the proposed bound gives $\alpha\sigma^2-E_x/N_0$, which may be
significantly larger for large $\sigma^2$. Moreover, returning to
(\ref{phase-est}), if $\sigma^2$ is large,
then $\tsig^2$ should be chosen large as well (otherwise the divergence
between the priors would be large), and then the terms containing
$e^{-\tsig^2/2}$ can be neglected. Under this approximation, it is easily seen
that the optimal choice of $\tsig^2$ is
\begin{equation}
\tsig^2=\frac{\sigma^2}{1-2\alpha\sigma^2},
\end{equation}
which when substituted back into (\ref{phase-est}), yields
\begin{equation}
\Lambda_{\mbox{\tiny B}}(\ct,\alpha)\ge
\frac{1}{2}\ln\frac{1}{1-2\alpha\sigma^2}-\frac{E_x}{N_0},
\end{equation}
which is finite only as long as $\alpha < \frac{1}{2\sigma^2}$, namely,
in this case, $\alpha_{\mbox{\tiny c}}$ is upper bounded by 
\begin{equation}
\alpha_{\mbox{\tiny c}}\le \frac{1}{2\sigma^2}.
\end{equation}
It turns out that this upper bound is tight, as it is
achieved at least by the trivial estimator $\ct\equiv 0$, whose estimation error,
$\epsilon=0-\Theta=-\Theta$ is indeed Gaussian with variance $\sigma^2$, by the model
assumption. In other words, for the model considered here,
\begin{equation}
\alpha_{\mbox{\tiny c}}= \frac{1}{2\sigma^2}.
\end{equation}
We observe that in this case of the non--linear signal model, the Gaussian tail of the pdf 
of the estimation error is due to the prior only, and
the observation set $\bY$ contributes nothing whatsoever to this Gaussian
tail. This is different from the Gaussian--linear model considered before, where we
found that $\alpha_{\mbox{\tiny c}}= \frac{1}{2\sigma^2}+\frac{E_s}{N_0}$,
which contains contributions of both the prior (the first term) and the
observation set (the second term).

\subsection{More General Priors}

So far, we have considered only Gaussian priors, for both $P$ and for $Q$.
This limits the framework to models where the support $\calA$ of the parameter
is the entire real line. On the one hand, for reasons that were discussed
above, it is convenient to work with a
Gaussian prior for $Q$, but on the other hand, if the parameter $\theta$, 
under the real model $P$, takes
values only in a finite interval $\cal$ (like in the case of a delay
parameter, for instance), then the divergence between the two priors
would be infinite and then our lower bound would be useless.
This motivates us to extend the scope to general priors whose support may not
necessarily be the entire real line.

Considering the case where the prior is $Q(\theta)$, and $Q(\by|\theta)$ is
according to (\ref{ref}), then under certain regularity conditions
(see, e.g., \cite{vantrees}), the Bayesian Cram\'er--Rao lower bound is given by
\begin{equation}
\bE_Q(\ct-\theta)^2\ge \frac{1}{I(Q)+2E_s/N_0}
\end{equation}
where
\begin{equation}
I(Q)=\bE_Q\{W^2(\Theta,\bY)\}=
\int_{\calA}\frac{[Q'(\theta)]^2}{Q(\theta)}\mbox{d}\theta .
\end{equation}
Given a general prior $P$, a convenient choice for $Q$, here indexed by
an auxiliary parameter $\beta > 0$, is of the form
\begin{equation}
Q_\beta(\theta)=\frac{P^\beta(\theta)}{Z(\beta)}~~~~~Z(\beta)=\int_{\calA}P^\beta(\theta)\mbox{d}\theta.
\end{equation}
Denoting $\phi(\beta)=\ln Z(\beta)$, the Kullback--Leibler divergence is
given by
\begin{equation}
D(Q_\beta\|P)=(\beta-1)\phi^\prime(\beta)-\phi(\beta).
\end{equation}
The idea is that one can now optimize the bound w.r.t.\ $\beta$ and $E_s$ (or
equivalently, $\lambda=E_s/N_0$), as the optimal signal $s(t)$ for a given $E_s$ is the same
as before. Of course, for a general $Q$, the calculation of $\bE_Q\{\Theta
x(t,\Theta)\}$ will have to be modified accordingly, as earlier we have calculated
it with a Gaussian prior. This quantity depends, of course, on $\beta$.

Consider, for example, the model of delay estimation, 
where under $P$, $x(t,\theta)=x(t-\theta)$, and assume that the derivative
$\dot{x}(t)$ has
finite energy. Suppose that under $Q$, the signal is $s(t-\theta)$, where 
$s(\cdot)$ is subjected to optimization of the bound.
In this case, the lower bound is of the form,
\begin{equation}
\Lambda_{\mbox{\tiny B}}(\ct,\alpha)
\ge\frac{\alpha}
{I(Q_\beta)+\frac{2}{N_0}\int_0^T[\dot{s}(t)]^2
\mbox{d}t}-
(\beta-1)\phi^\prime(\beta)+\phi(\beta)-
\frac{1}{N_0}\int_0^T[x(t)-s(t)]^2\mbox{d}t.
\end{equation}
As can be seen,
the optimization over the signal $s(\cdot)$ involves a tradeoff between
the energy of its derivative and the Euclidean distance between $s(t)$ and $x(t)$.
In particular,
\begin{eqnarray}
\Lambda_{\mbox{\tiny B}}(\ct,\alpha)&\ge&
\sup_{s(\cdot),\beta}\left\{\frac{\alpha}
{I(Q_\beta)+\frac{2}{N_0}\int_0^T\dot{s}^2(t)
\mbox{d}t}
-(\beta-1)\phi^\prime(\beta)+\phi(\beta)\right.\nonumber\\
& &\left.-\frac{1}{N_0}\int_0^T[x(t)-s(t)]^2\right\}\nonumber\\
&=&\sup_{\tilde{E},\beta}\left[\frac{\alpha}
{I(Q_\beta)+2\tilde{E}/N_0}
-(\beta-1)\phi^\prime(\beta)+\phi(\beta)-\right.\nonumber\\
& &\left.\frac{1}{N_0}\inf
\left\{\int_0^T[x(t)-s(t)]^2\mbox{d}t:~\int_0^T\dot{s}^2(t)\mbox{d}t=
\tilde{E}\right\}\right].
\end{eqnarray}
The inner minimization can be cast as a Lagrangian minimization
\begin{equation}
\inf_{s(\cdot)}\left\{\int_0^T\dot{s}^2(t)\mbox{d}t+\lambda
\int_0^T[x(t)-s(t)]^2\mbox{d}t\right\},
\end{equation}
which can easily be solved using calculus of variations.
In particular, suppose that $s_*(t)$ is the optimal signal and consider
a perturbation $s(t)=s_*(t)+\epsilon r(t)$ for an arbitrary signal $r(t)$.
On substituting this form back into the Lagrangian and requiring
that the derivative vanishes at $\epsilon=0$ for every $r(t)$, we obtain the
equation,
\begin{equation} 
\int_0^T\{\dot{r}(t)\dot{s}_*(t)+\lambda r(t)[s_*(t)-x(t)]\}\mbox{d}t=0,
\end{equation}
or equivalently (using integration by parts),
\begin{equation} 
r(T)\dot{s}_*(T)-r(0)\dot{s}_*(0)+\int_0^Tr(t)\cdot\{-\ddot{s}_*(t)+\lambda[s_*(t)-x(t)]\}\mbox{d}t=0,
\end{equation}
for every $\{r(t),~0\le t\le T\}$,
which means that $s_*(t)$ is obtained from $x(t)$ by solving the second order
linear differential equation,
\begin{equation}
s_*(t)-\frac{\ddot{s}_*(t)}{\lambda}=x(t)~~~~~~\dot{s}_*(0)=\dot{s}_*(T)=0.
\end{equation}
As an example, let
\begin{equation}
x(t)=\sqrt{\frac{2E_x}{3T}}\cdot[1-\cos(\omega_0 t)],
\end{equation}
where $\omega_0$ is a known parameter.
Then the solution to the above differential equation is
given by
\begin{equation}
s_*(t)=\sqrt{\frac{2E_x}{3T}}\cdot\left[1-\frac{\lambda}{\lambda+\omega_0^2}\cdot\cos(\omega_0
t)\right].
\end{equation}
Now, denoting $\nu=\lambda/{(\lambda+\omega_0^2)}$, we have
\begin{equation}
\frac{1}{N_0}\int_0^T[s_*(t)-x(t)]^2\mbox{d}t
=\frac{E_x}{3N_0}\cdot(1-\nu)^2
\end{equation}
and 
\begin{equation}
\int_0^T\dot{s}_*^2(t)\mbox{d}t=\frac{1}{3}E_x\omega_0^2\cdot\nu^2.
\end{equation}
Thus, our lower bound can be presented as
\begin{equation}
\Lambda_{\mbox{\tiny B}}(\ct,\alpha)\ge
\sup_{0\le\nu\le 1,~\beta> 0}\left[\frac{\alpha}
{I(Q_\beta)+2\nu^2\omega_0^2E_x/3N_0}
-(\beta-1)\phi^\prime(\beta)+\phi(\beta)-\frac{E_x}{3N_0}\cdot(1-\nu)^2\right].
\end{equation}
Note that by selecting $\nu=0$, this can be further lower bounded by
\begin{eqnarray}
\Lambda_{\mbox{\tiny B}}(\ct,\alpha)&\ge&
\sup_{\beta\ge 0}\left[\frac{\alpha}
{I(Q_\beta)}
-(\beta-1)\phi^\prime(\beta)+\phi(\beta)-\frac{E_x}{3N_0}\right]\\
&=&\sup_{\beta\ge
0}\left[\frac{\alpha-I(Q_\beta)[(\beta-1)\phi^\prime(\beta)-\phi(\beta)]}
{I(Q_\beta)}
-\frac{E_x}{3N_0}\right],
\end{eqnarray}
which is finite only as long as $\alpha < \alpha_{\mbox{\tiny c}}$, 
where $\alpha_{\mbox{\tiny c}}$ can be evaluated (or at least,
upper bounded) by $\lim_{\beta\to
\beta_0}I(Q_\beta)[(\beta-1)\phi^\prime(\beta)-\phi(\beta)]$,
where $\beta_0$ is such that $\lim_{\beta\to\beta_0} I(Q_\beta)=0$
(for example, if the prior is Gaussian, then $\beta_0=0$ and correspondingly,
$\alpha_{\mbox{\tiny c}}=\frac{1}{2\sigma^2}$).
If no such $\beta_0$ exists, 
then the upper bound to $\alpha_{\mbox{\tiny c}}$ is infinity. 

The case of frequency estimation, where
$x(t,\theta)=a(t)\sin(\theta
t)+b(t)\cos(\theta t)$ can be handled similarly. In that case, 
the optimum auxiliary signal turns out to be
$s(t,\theta)=\lambda x(t,\theta)/(t^2+\lambda)$, where $\lambda$ is again an
auxiliary parameter that controls the tradeoff between the energy 
of the derivative of $s(t,\theta)$ and its distance to $x(t,\theta)$.

\subsection{Risk--Sensitive Bounds Based on Other MSE Lower Bounds}

So far, our lower bounds on the risk--sensitive cost function were based
solely on
the Bayesian Cram\'er--Rao lower bound as an MSE lower bound for the reference
model. Of course, other MSE lower bounds may be used as well, and one of the
best such bounds is the Weiss--Weinstein bound \cite{vantreesbell}, \cite{weissphd}.
In this short subsection, we demonstrate its
usefulness in the problem of delay estimation for a rectangular pulse, where the delay parameter has a
uniform prior over a certain interval.

For the case where the underlying signal is a
rectangular pulse of width $\tau$, the MMSE of its delay was shown
in \cite{weissphd} to be lower bounded by $0.324\tau^2/\gamma^2$,
where $\gamma=E_x/N_0$ (and in the literature, there are other bounds with different constants,
but the same behavior of being proportional to $\tau^2/\gamma^2$).
Considering an auxiliary rectangular pulse (under $Q$) with the
same energy, but duration $\tilde{\tau}\ge\tau$, we have
\begin{equation}
\Lambda_{\mbox{\tiny B}}(\ct,\alpha)\ge
\frac{0.324\tilde{\tau}^2}{\gamma^2}-2\gamma\left(1-\sqrt{\frac{\tau}{\tilde{\tau}}}\right),
\end{equation}
where the second term is due to the divergence between $Q$ and $P$ (which is
proportional to the Euclidean distance between the two rectangular pulses), and 
where we let the prior over auxiliary prior be the same as
the original one.
Optimizing over $\tilde{\tau}$, we obtain
\begin{equation}
\tilde{\tau}^*=1.1895\left(\frac{\gamma^6\tau}{\alpha^2}\right)^{0.2},
\end{equation}
and the assumption $\tilde{\tau}^*\ge\tau$ limits the applicability of this
result to the case $\gamma\ge 0.8654(\alpha\tau^2)^{1/3}$. On substituting
$\tilde{\tau}^*$ back into the above lower bound, we obtain
\begin{equation}
\Lambda_{\mbox{\tiny B}}(\ct,\alpha)\ge
2.2922(\alpha\gamma^2\tau^2)^{0.2}-2\gamma,
\end{equation}
which is non--trivial (non--negative) as long as 
\begin{equation}
\gamma < 1.2552
(\alpha\tau^2)^{1/3}. 
\end{equation}
The bound seems to be monotonically increasing in $\gamma$
in some range of small $\gamma$, but this range is outside the range of
applicability, 
\begin{equation}
\gamma\ge 0.8654(\alpha\tau^2)^{1/3}.
\end{equation}

\subsection{The Logarithmic Probability Comparison Bound}

As mentioned earlier in this section, for the linear--Gaussian 
model (\ref{ref}), the conditional mean
estimator (\ref{condmean}) minimizes all exponential moments of the squared error, and
the performance is
\begin{equation}
\Lambda_{\mbox{\tiny B}}(\ct,\alpha)=
\frac{1}{2}\ln\frac{1}{1-\alpha/
\alpha_{\mbox{\tiny c}}},
\end{equation}
where here, once again, $\alpha_{\mbox{\tiny
c}}=\frac{E_s}{N_0}+\frac{1}{2\sigma^2}$.
Returning to the model $P$, where $\Theta\sim\calN(0,\sigma^2)$ and
$y(t)=x(t,\theta)+n(t)$, and letting $Q$ be again defined by
$\Theta\sim\calN(0,\tsig^2)$ and
$y(t)=\theta s(t)+n(t)$, in this section, we use a more general family of
inequalities, where the Kullback--Leibler divergence is replaced
by the R\'enyi divergence. This family of inequalities induces a class of bounds 
referred to as the {\it logarithmic probability comparison bound} (LPCB),
\cite{ACD15}, \cite{AM15}.

\begin{figure}[h!t!b!]
\centering
\includegraphics[width=5.5cm, height=5.5cm]{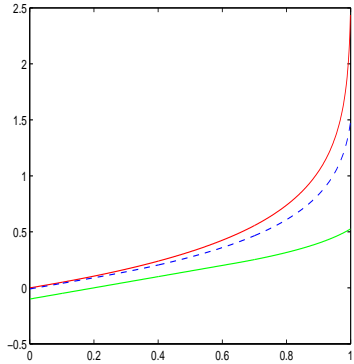}
\caption{\small Graphs
of the r.h.s.\ of eq.\ (\ref{lastbound}) as a function of $\alpha$. Here, $\sigma^2=1/2$,
$E_s=q=0$ (hence $\tilde{\alpha}_{\mbox{\tiny c}}=1$) and
the SNR takes the values: $E_x/N_0=0.001$ (red),
$E_x/N_0=0.01$ (blue), and
$E_x/N_0=0.1$ (green).}
\label{lowerbound}
\end{figure}

The idea behind the LPCB bound is as follows. For a given $a > 1$, let
us define the R\'enyi divergence as
\begin{equation}
D_a(Q\|P)=\frac{1}{a(a-1)}\ln\int\left(\frac{\mbox{d}Q}{\mbox{d}P}\right)^a\mbox{d}P.
\end{equation}
Then, for a given random variable $X$, governed by either $P$ or $Q$,
the underlying family of inequalities \cite{ACD15} is as follows:
\begin{equation}
\ln \bE_P e^{aX} \ge \frac{a}{a-1}\ln\bE_Q e^{(a-1)X}-aD_a(Q\|P).
\end{equation}
Note that the inequality $\ln E_P e^X \ge \bE_Q X-D(Q\|P)$, that we have used
in all previous parts of the paper, is just a special case of the last
inequality, where $a\downarrow 1$.
Applying this more general 
inequality with $X=\beta(\ct-\theta)^2$ ($\beta > 0$, a given
constant) and defining $\alpha=a\beta$, we obtain the
following inequality for every $\alpha > \beta > 0$:
\begin{eqnarray}
\Lambda_{\mbox{\tiny B}}(\ct,\alpha)&\ge&
\frac{\alpha}{\alpha-\beta}\Lambda_{\mbox{\tiny B}}(\ct,\alpha-\beta)
-\frac{\alpha}{\beta}D_{\alpha/\beta}(Q\|P)\\
&\ge&-\frac{\alpha}{2(\alpha-\beta)}\ln\left(1-\frac{\alpha-\beta}
{\tilde{\alpha}_{\mbox{\tiny c}}}\right)-
\frac{\alpha}{\beta}D_{\alpha/\beta}(Q\|P)\\
&=&-\frac{\alpha}{2(\alpha-\beta)}\ln\left(1-\frac{\alpha-\beta}
{\tilde{\alpha}_{\mbox{\tiny c}}}\right)-
\frac{\beta}{\alpha-\beta}\ln\int
\left(\frac{\mbox{d}Q}{\mbox{d}P}\right)^{\alpha/\beta}\mbox{d}P,
\end{eqnarray}
where $\tilde{\alpha}_{\mbox{\tiny c}}=\frac{1}{2\tsig^2}+\frac{E_s}{N_0}$.
Now,
\begin{eqnarray}
& &\int\left(\frac{\mbox{d}Q}{\mbox{d}P}\right)^a\mbox{d}P\nonumber\\
&=&\bE_P\left\{\left[\frac{(2\pi\tsig^2)^{-1/2}\exp\{-\Theta^2/2\tsig^2\}}
{(2\pi\sigma^2)^{-1/2}\exp\{-\Theta^2/2\sigma^2\}}\right]^a
\exp\left[\frac{a(a-1)}{N_0}\int_0^T[x(t,\Theta)-\Theta
s(t)]^2\mbox{d}t\right]\right\}\nonumber\\
&=&\left(\frac{\sigma^2}{\tsig^2}\right)^{a/2}
\bE_P\left\{\exp\left[a\left(\frac{1}{2\sigma^2}-\frac{1}{2\tsig^2}\right)\Theta^2
+\frac{a(a-1)}{N_0}\int_0^T[x(t,\Theta)-\Theta
s(t)]^2\mbox{d}t\right]\right\}\nonumber\\
&=&\left(\frac{\sigma^2}{\tsig^2}\right)^{a/2}
\bE_P\left\{\exp\left(a\left[\frac{1}{2\sigma^2}-\frac{1}{2\tsig^2}\right]\Theta^2
+\frac{a(a-1)}{N_0}\left[E_x+\Theta^2E_s-2\Theta\int_0^Ts(t)x(t,\Theta)
\mbox{d}t\right]\right)\right\}.\nonumber
\end{eqnarray}
Let us now assume that $\int_0^Tx(t,\theta)\mbox{d}t$ is a constant,
independent of $\theta$ (which is as reasonable as assuming that the energy is
independent of $\theta$), and denote this constant by
$q$. Then, letting
$s(t)\equiv\sqrt{E_s/T}$ (DC signal), we get
\begin{eqnarray}
& &\int\left(\frac{\mbox{d}Q}{\mbox{d}P}\right)^a\mbox{d}P\nonumber\\
&=&\left(\frac{\sigma^2}{\tsig^2}\right)^{a/2}
\bE_P\left\{\exp\left(a\left[\frac{1}{2\sigma^2}-\frac{1}{2\tsig^2}\right]\Theta^2
+\frac{a(a-1)}{N_0}\left[E_x+\Theta^2E_s-2\Theta q\sqrt{\frac{E_s}{T}}
\right]\right)\right\}\nonumber\\
&=&\left(\frac{\sigma^2}{\tsig^2}\right)^{a/2}
\exp\left\{\frac{a(a-1)E_x}{N_0}\right\}\cdot
\bE_P\exp\left\{\left[\frac{a}{2\sigma^2}-\frac{a}{2\tsig^2}+\frac{a(a-1)E_s}{N_0}\right]\Theta^2
-\frac{2a(a-1)q}{N_0}\sqrt{\frac{E_s}{T}}\cdot\Theta\right\}.\nonumber
\end{eqnarray}
Now,
\begin{equation}
\bE_P\exp\{A\Theta^2-B\Theta\}=\frac{\exp\left\{\frac{B^2\sigma^2}{2(1-2A
\sigma^2)}\right\}}{\sqrt{1-2A\sigma^2}},
\end{equation}
where in our case,
\begin{equation}
A=\frac{a}{2\sigma^2}-\frac{a}{2\tsig^2}+\frac{a(a-1)E_s}{N_0}
=\frac{\alpha}{2\beta\sigma^2}-\frac{\alpha}{2\beta\tsig^2}+\frac{\alpha(\alpha-\beta)E_s}{\beta^2N_0}
\end{equation}
and
\begin{equation}
B=\frac{2a(a-1)q}{N_0}\sqrt{\frac{E_s}{T}}
=\frac{2\alpha(\alpha-\beta)q}{\beta^2N_0}\sqrt{\frac{E_s}{T}}
\end{equation}
Note that as an alternative to the assumption that $\int_0^T x(t,\theta)\mbox{d}t=q$ for
all $\theta$, one might assume that $s(t)$ is orthogonal to $x(t,\theta)$ for
all $\theta$, which is equivalent to $q=0$, for which $B=0$.
It follows that
\begin{eqnarray}
aD_a(Q\|P)&=&\frac{1}{a-1}\ln\int\left(\frac{\mbox{d}Q}{\mbox{d}P}\right)^a\mbox{d}P\\
&=&\frac{a}{2(a-1)}\ln\frac{\sigma^2}{\tsig^2}+\frac{aE_x}{N_0}+\frac{B^2\sigma^2}{2(a-1)(1-2A
\sigma^2)}-\frac{1}{2(a-1)}\ln(1-2A\sigma^2).
\end{eqnarray}
Thus, the lower bound becomes
\begin{eqnarray}
\Lambda_{\mbox{\tiny B}}(\ct,\alpha)
&\ge&-\frac{\alpha}{2(\alpha-\beta)}\ln\left(1-\frac{\alpha-\beta}{\tilde{\alpha}_{\mbox{\tiny
c}}}\right)-\frac{\alpha}{2(\alpha-\beta)}\ln\frac{\sigma^2}{\tsig^2}
-\frac{\alpha E_x}{\beta N_0}\nonumber\\
& & -\frac{\beta B^2\sigma^2}{2(\alpha-\beta)(1-2A
\sigma^2)}+\frac{\beta}{2(\alpha-\beta)}\ln(1-2A\sigma^2),
\end{eqnarray}
where we have the freedom to maximize over the variables $\beta$ (in the range
$\beta\in(0,\alpha)$), $E_s\ge 0$ and
$\tsig^2> 0$. Choosing, for example, $\tsig^2=\sigma^2$ and $E_s\to 0$ (which
yield $A\to 0$ and $B\to 0$), we
get
\begin{equation}
\label{lastbound}
\Lambda_{\mbox{\tiny B}}(\ct,\alpha)
\ge \sup_{0<\beta<\alpha}\left\{\frac{\alpha}{2(\alpha-\beta)}
\ln\left[\frac{1}{1-2\sigma^2(\alpha-\beta)}\right]
-\frac{\alpha E_x}{\beta N_0}\right\},
\end{equation}
which, for $\beta\to 0$, limits $\alpha_{\mbox{\tiny c}}$ to be no more than
$1/2\sigma^2$ 
(which is again, obviously achievable, e.g., by $\ct\equiv
0$). To see why this is true, 
suppose conversely, that $\alpha=\frac{1+\epsilon}{2\sigma^2}$ for some
arbitrarily small $\epsilon > 0$. Then by choosing
$\beta=\frac{\epsilon}{2\sigma^2}$, 
the above lower bound becomes infinite. 
Some graphs of the last bound are depicted in Fig.\ \ref{lowerbound}.

\subsubsection*{Iterated LPCB}

Let $\alpha$ be given and let $\beta_1,\beta_2,\ldots,\beta_k$ be positive
numbers such that $\sum_{i=1}^k\beta_i < \alpha$. Let $P_1$, $P_2,\ldots, P_k$ be a
corresponding a sequence of probability measures with $P_1\equiv P$ and
$P_k\equiv Q$, and let $\bE_i$ denote the expectation operator w.r.t.\ $P_i$,
$i=1,2,\ldots,k$.
Now consider the following chain of inequalities:
\begin{eqnarray}
\Lambda_{\mbox{\tiny B}}(\ct,\alpha)&\ge&
\frac{\alpha}{\alpha-\beta_1}\ln\bE_2
e^{(\alpha-\beta_1)(\ct-\theta)^2}-\frac{\alpha}{\beta_1}D_{\alpha/\beta_1}(P_2\|P_1)\\
&\ge&
\frac{\alpha}{\alpha-\beta_1}\left[\frac{\alpha-\beta_1}{\alpha-\beta_1-\beta_2}\ln\bE_3
e^{(\alpha-\beta_1-\beta_2)(\ct-\theta)^2}-\frac{\alpha-\beta_1}{\beta_2}
D_{(\alpha-\beta_1)/\beta_2}(P_3\|P_2)\right]-\nonumber\\
& &\frac{\alpha}{\beta_1}D_{\alpha/\beta_1}(P_2\|P_1)\\
&=&
\frac{\alpha}{\alpha-\beta_1-\beta_2}\ln\bE_3
e^{(\alpha-\beta_1-\beta_2)(\ct-\theta)^2}-\frac{\alpha}{\beta_2}
D_{(\alpha-\beta_1)/\beta_2}(P_3\|P_2)-\nonumber\\
& &\frac{\alpha}{\beta_1}D_{\alpha/\beta_1}(P_2\|P_1)\\
&\ge&\frac{\alpha}{\alpha-\sum_{i=1}^k\beta_i}\ln\bE_k
\exp\left\{\left(\alpha-\sum_{i=1}^k\beta_i\right)(\ct-\theta)^2\right\}-\nonumber\\
& &\alpha\sum_{i=1}^{k-1}\frac{1}{\beta_i}
D_{(\alpha-\sum_{j=1}^{i-1}\beta_j)/\beta_i}(P_{i+1}\|P_i).
\end{eqnarray}
We now have the freedom to choose the
parameters $k$, $\beta_1,\ldots,\beta_k$ and $P_2, P_3,\ldots,P_k$.
The original inequality we have been using (\ref{genericb}) is obtained as a special case where
$\beta_i\downarrow 0$ for $i=1,2,\ldots,k-2,k$, $\beta_{k-1}\uparrow \alpha$,
and $P_2=P_3=\ldots=P_{k-1}=P$ (one can also choose, of course, $k=2$ and
$\beta_1\uparrow\alpha$, $\beta_2\downarrow 0$).
The ordinary (one--step) LPBC bound is obtained by choosing
$P_2=P_3=\ldots=P_{k-1}=Q$, $\beta_1=\beta < \alpha$ and
$\beta_2=\beta_3=\ldots=\beta_k\downarrow 0$
(one can also choose, of course, $k=2$ and
$\beta_1=\beta$, $\beta_2\to 0$). The iterated LPCB can be useful in
situations where it is easier to derive (or to bound) the R\'enyi divergences
via certain intermediate reference models rather than going directly from the
underlying model to the ultimate reference model (see, e.g., the last part of
Subsection 4.1 in \cite{AM15}).

\subsection{Phase Transitions}

In this subsection, unlike all other parts of this paper, our focus is not
quite only on lower bounds and their tightness, but rather on another aspect
of the risk--sensitive cost function, and this is that in some cases, this
cost function may exhibit phases transitions, even in rather seemingly simple
and innocent parametric models. We have already seen that in many situations
where $\theta$ is unbounded, there might be a critical value of the
risk--sensitive factor, $\alpha_{\mbox{\tiny c}}$ beyond which this cost
function diverges. This is certainly a very sharp (first--order) phase
transition. However, in some cases there might be additional phase
transitions, due to possible analogies with certain models in statistical
mechanics, and the purpose of this subsection is to demonstrate this fact.

\begin{figure}[h!t!b!]
\centering
\includegraphics[width=5.5cm, height=5.5cm]{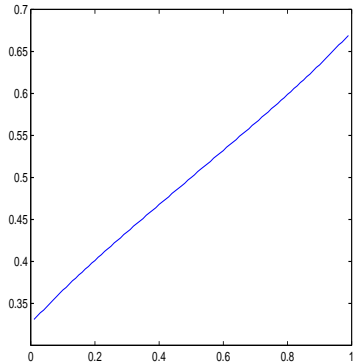}
\caption{\small Graph of $\ct(q)$. Note that the range of $\ct(q)$, which is
$[0.331,0.669]$, is much smaller
than the domain of $q$, which is $[0,1]$. The estimator avoids extreme values
in order not to incur large errors in case the true $\theta$ is in the other
extreme. This is due
to its risk--sensitive nature.}
\label{est}
\end{figure}

Consider the example where $\Theta\sim\mbox{Unif}[0,1]$, $\by\in\{0,1\}^n$
and $P(\by|\theta)$ is the Bernoulli distribution,
\begin{equation}
P(\by|\theta)=\theta^{nq}(1-\theta)^{n(1-q)},
\end{equation}
where $nq=\sum_{i=1}^ny_i$ is the number of ones.
We would like to investigate the behavior of the risk--sensitive cost function,
and our focus will be on the
asymptotic behavior at the exponential scale, where $\alpha$ will be assumed
to be growing linearly with $n$, that is, we take $\alpha=a\cdot n$ for some
fixed $a > 0$.
We will study the asymptotic exponent of the optimal estimator as a function of $\theta$ and $a$.
The optimal Bayesian performance is given by 
\begin{eqnarray}
\min_{\ct}\bE\exp\left\{\alpha n(\ct-\theta)^2\right\}&=&
\sum_{\by}P(\by)\min_t\int_0^1 \exp\left\{\alpha
n(t-\theta)^2\right\}P(\theta|\by)\mbox{d}\theta\\
&\exe&\sum_{\by}P(\by)\min_{t\in[0,1]}\max_{\theta\in[0,1]} \exp\left\{n
[a (t-\theta)^2-D(q\|\theta)]\right\}\\
&\exe&\exp\left\{n\cdot\max_{q\in[0,1]}\min_{t\in[0,1]}\max_{\theta\in[0,1]}
\left[a (t-\theta)^2-D(q\|\theta)\right]\right\},
\end{eqnarray}
where $\exe$ denotes equality in the exponential scale, and
where $D(q\|\theta)$ designates the binary divergence function,
\begin{equation}
D(q\|\theta)=q\ln\frac{q}{\theta}+(1-q)\ln\frac{1-q}{1-\theta}.
\end{equation}
The (asymptotically) optimal estimator is then given by
\begin{equation}
\ct=\arg\min_{t\in[0,1]}\max_{\theta\in[0,1]}\left[a
(t-\theta)^2-D(q\|\theta)\right],
\end{equation}
which depends on $\by$ solely via $q$. This estimator is depicted in Fig.\
\ref{est} for
$a=10$.

\begin{figure}[h!t!b!]
\centering
\includegraphics[width=5.5cm, height=5.5cm]{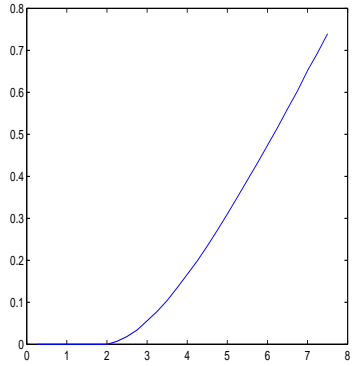}
\caption{\small Graph of $E(a)$.}
\label{errorexp}
\end{figure}

Obviously,
\begin{eqnarray}
E(a)&\dfn&\max_{q\in[0,1]}\min_{t\in[0,1]}\max_{\theta\in[0,1]}
\left[a (t-\theta)^2-D(q\|\theta)\right]\\
&\ge&\max_{q\in[0,1]}\min_{t\in[0,1]} a (t-q)^2\\
&\ge& 0.
\end{eqnarray}
On the other hand,
note that by the Pinsker inequality \cite{CK81}, 
$D(q\|\theta)\ge 2(q-\theta)^2$, and so,
for $a\in[0,2]$
\begin{eqnarray}
E(a)&\le&\max_{q\in[0,1]}\min_{t\in[0,1]}\max_{\theta\in[0,1]}
\left[a (t-\theta)^2-2(q-\theta)^2\right]\\
&\le&\max_{q\in[0,1]}\max_{\theta\in[0,1]}
\left[a (q-\theta)^2-2(q-\theta)^2\right]\\
&=&\max_{q\in[0,1]}\max_{\theta\in[0,1]}
(a-2)(q-\theta)^2\\
&=&0,
\end{eqnarray}
and so, it follows that for $a\in[0,2]$, $E(a)=0$
and it is achieved by
$\ct=q$. The following graph (Fig.\ \ref{errorexp}) displays $E(a)$
and it suggests
that $a=2$ is indeed the maximum value of $a$ for which
it still vanishes. Thus, we see a phase transition at $a=2$.

\begin{figure}[ht]
\hspace*{3cm}\input{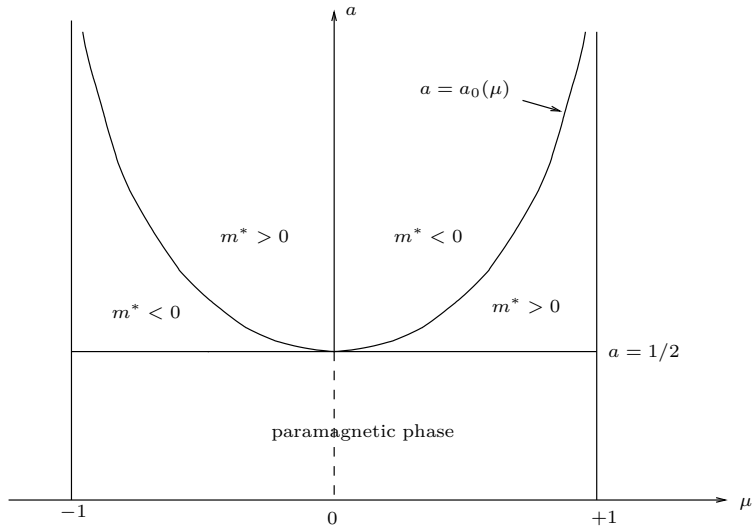}
\caption{\small Phase diagram in the plane of $(\mu,\alpha)$.}
\label{mm1}
\end{figure}

We have seen that the estimator $\ct=q$ is asymptotically (exponentially) 
optimal for every $a\le 2$.
We next analyze the risk--sensitive cost function for this estimator on the
exponential scale, and as said, we demonstrate that it also exhibits phase
transitions. As already mentioned, it turns
out that in some cases, the expression of the exponential moment is analogous
to that of a partition function of a certain physical model, in particular, a
model of magnetic
spins with interactions, which may exhibit phase transitions. 
To facilitate
the presentation of the analogy with statistical mechanics, we now slightly
modify our notation.
Consider the case where $\bX$ is a binary vector whose components take
on values in $\calX=\{-1,+1\}$, and which is governed by a binary memoryless
source $P_\mu$ with probabilities
$\mbox{Pr}\{X_i=+1\}=1-\mbox{Pr}\{X_i=-1\}=(1+\mu)/2$, ($\mu$ designating the
expected `magnetization' of each binary spin $X_i$).
The probability of $\bx$ under $P_\mu$ is thus easily
shown to be given by
\begin{equation}
P_\mu(\bx)=\left(\frac{1+\mu}{2}\right)^{(n+\sum_ix_i)/2}\cdot
\left(\frac{1-\mu}{2}\right)^{(n-\sum_ix_i)/2}
=\left(\frac{1-\mu^2}{4}\right)^{n/2}\cdot\left(\frac{1+\mu}{1-\mu}\right)^{\sum_ix_i/2}.
\end{equation}
Consider the estimation of the parameter $\mu$ by the ML estimator
\begin{equation}
\hat{\mu}=\frac{1}{n}\sum_{i=1}^nx_i.
\end{equation}
Now,
\begin{eqnarray}
\bE_\mu\exp\{a n(\hat{\mu}-\mu)^2\}&=&
\left(\frac{1-\mu^2}{4}\right)^{n/2}e^{na\mu^2}
\sum_{\bx}\left(\frac{1+\mu}{1-\mu}\right)^{\sum_ix_i/2}
\exp\left\{\frac{a}{n}\left(\sum_ix_i\right)^2
-2a\mu\sum_ix_i\right\}\nonumber\\
&=&\left(\frac{1-\mu^2}{4}\right)^{n/2} e^{na\mu^2}
\sum_{\bx}\exp\left\{\left(\frac{1}{2}\ln\frac{1+\mu}{1-\mu}-2a\mu\right)\sum_ix_i+
\frac{a}{n}\left(\sum_ix_i\right)^2\right\}.\nonumber
\end{eqnarray}
The last summation over $\{\bx\}$ is exactly the partition function pertaining
to the {\it Curie--Weiss model} of spin arrays in statistical mechanics (see,
e.g.,
\cite[Subsection 2.5.2]{MM09}), where
the magnetic field is given by
\begin{equation}
B=\frac{1}{2}\ln\frac{1+\mu}{1-\mu}-2a\mu
\end{equation}
and the coupling coefficient for every pair of spins is $J=2a$.
It is well known that this model exhibits phase transitions pertaining to
spontaneous magnetization below a certain critical temperature. In particular,
using the method of types \cite{CK81}, this partition function can be
asymptotically evaluated
as being of the exponential order of
\begin{equation}
\exp\left\{n\cdot\max_{|m|\le
1}\left[h_2\left(\frac{1+m}{2}\right)+Bm+\frac{J}{2}\cdot
m^2\right]\right\},
\end{equation}
where $h_2(u)=-u\ln u-(1-u)\ln(1-u)$ is the binary entropy function, which stands for the
exponential order of the number of configurations $\{\bx\}$ with a given
value of $m=\frac{1}{n}\sum_ix_i=\hat{\mu}$. This expression
is clearly dominated by a
value of $m$ (the dominant magnetization $m^*$)
which maximizes the expression in the square brackets, i.e., it solves the
equation
\begin{equation}
m=\tanh(Jm+B),
\end{equation}
or in our variables,
\begin{equation}
m=\tanh\left(2a m+\frac{1}{2}\ln\frac{1+\mu}{1-\mu}-2a\mu\right).
\end{equation}
For $a < 1/2$, there is only one solution and there is no spontaneous
magnetization (paramagnetic phase). For $a > 1/2$, however, there are
three solutions, and only one of them dominates the partition function,
depending on the sign of $B$, or equivalently, on whether $a >
a_0(\mu)\dfn\frac{1}{4\mu}\ln\frac{1+\mu}{1-\mu}$ or $a <
a_0(\mu)$ and according to the sign of $\mu$. Accordingly, there are five
different phases in the plane spanned by $a$ and $\mu$. The paramagnetic
phase $a < 1/2$, the phases $\{\mu > 0,~1/2 < a < a_0(\mu)\}$
and $\{\mu < 0,~ a > a_0(\mu)\}$, where the dominant magnetization
$m$ is positive, and the two complementary phases,
$\{\mu < 0,~1/2 < a < a_0(\mu)\}$ and
and $\{\mu > 0,~ a > a_0(\mu)\}$, where the dominant magnetization
is negative. Thus, there is a multi-critical point
where the boundaries of all five phases meet,
which is the point $(\mu,a)=(0,1/2)$. The phase diagram is depicted in Fig.\
\ref{mm1}.

This analysis can easily be extended to the
case of a
general parametric family of discrete memoryless source (DMSs),
$\{P(\cdot|\theta),~\theta\in\calA\}$, where
$\theta$ is a parameter vector, and one may replace the square error function
by a general loss function $L(\theta,\ct)$. The class
$\{P(\cdot|\theta),~\theta\in\calA\}$ may not necessarily span the entire simplex of
DMSs of the given alphabet size. 

\section{The Non--Bayesian Regime}

Many of the ideas raised in the previous section may have 
analogues in the non--Bayesian regime, but we will not derive all of
those analogous derivations herein. We will touch only a few important
points in this short section.

\subsection{The Linear Signal Model}

Consider again the signal model,
\begin{equation}
\label{signalinoise}
y(t)=x(t,\theta)+n(t),~~~~0\le t\le T
\end{equation}
where $n(t)$ is Gaussian white noise with two--sided spectral density $N_0/2$,
as before.
Consider, for example the case, $x(t,\theta)=\theta\cdot s(t)$, where
$\{s(t)\}$ has energy $E_s$. Then, if we take $L_{\mbox{\tiny NB}}(\ttt)$ to
be the Cram\'er--Rao bound, then it is 
given by $\frac{N_0}{2E_s}$ independent of $\ttt$, and
so, according to part 2 of Theorem 1, for any unbiased estimator,
\begin{eqnarray}
\Lambda_{\mbox{\tiny NB}}(\ct,\theta,\alpha)&\ge&\sup_{\ttt}\left\{\frac{\alpha
N_0}{2E_s}+\left(\alpha-\frac{E_s}{N_0}\right)(\ttt-\theta)^2\right\}\\
&=&\left\{\begin{array}{cc}
\frac{\alpha N_0}{2E_s} & \alpha \le \frac{E_s}{N_0}\\
\infty & \alpha > \frac{E_s}{N_0}\end{array}\right.
\end{eqnarray}
namely, $\alpha_{\mbox{\tiny c}}\le E_s/N_0$.
This means that no unbiased estimator has an estimation error $\epsilon$
with a tail that decays faster than $e^{-\epsilon^2E_s/N_0}$.
For the maximum likelihood (ML) estimator,
\begin{equation}
\ct_{\mbox{\tiny ML}}=\frac{1}{E_s}\int_0^Ty(t)s(t)\mbox{d}t=
\theta+\frac{1}{E_s}\int_0^Tn(t)s(t)\mbox{d}t
\end{equation}
and so, the estimation error
\begin{equation}
\epsilon= \ct-\theta=\frac{1}{E_s}\int_0^Tn(t)s(t)\mbox{d}t
\end{equation}
is Gaussian, zero--mean, with variance $\frac{N_0}{2E_s}$, which attains the
upper bound on $\alpha_{\mbox{\tiny c}}$. This means that
$E_s/N_0$ is the exact value of $\alpha_{\mbox{\tiny c}}$ and not merely an
upper bound. While in the Bayesian regime, we have for a similar model,
$\alpha_{\mbox{\tiny c}}=\frac{1}{2\sigma^2}+\frac{E_s}{N_0}$, here we do not
have the contribution of the prior, and the remaining term is just
$\frac{E_s}{N_0}$. It is interesting to note that we obtained a tight result
on $\alpha_{\mbox{\tiny c}}$, attained by the ML estimator, even though the bound
on $\Lambda_{\mbox{\tiny NB}}(\theta)$ is not attained by the ML estimator, as
\begin{equation}
\Lambda_{\mbox{\tiny NB}}(\ct_{\mbox{\tiny ML}},\theta,\alpha)
=-\frac{1}{2}\ln\left(1-\frac{\alpha
N_0}{E_s}\right).
\end{equation}
The ML estimator achieves, however, the bound
asymptotically when $\alpha N_0/E_s$ is very small, which means, either very
small $\alpha$ or very large signal--to--noise ratio, or both.

\subsection{The Vector Linear Signal Model}

The above derivation extends to the vector case quite easily. Let
$x(t,\theta)=\sum_{i=1}^k\theta_is_i(t)$, where $\{s_i(t)\}$ all have the same
energy, $E_s$ and let $\Gamma$ be the $k\times k$ matrix of correlations
with entries given by
\begin{equation}
\gamma_{ij}=\frac{1}{E_s}\int_0^Ts_i(t)s_j(t)\mbox{d}t.
\end{equation}
In this case, $D(\ttt\|\theta)=(\ttt-\theta)^T\Gamma(\ttt-\theta)$,
the Cram\'er--Rao lower bound is $N_0\Gamma^{-1}/2E_s$, and so, we get
\begin{eqnarray}
\ln\bE_\theta\exp\left\{[\alpha^T(\ct-\theta)]^2\right\}&\ge&
\frac{N_0\alpha^T\Gamma^{-1}\alpha}{2E_s}+\sup_{\ttt}
(\ttt-\theta)^T\left(\alpha\alpha^T-\frac{E_s}{N_0}\Gamma\right)(\ttt-\theta)\\
&=&\left\{\begin{array}{cc}
\frac{N_0\alpha^T\Gamma^{-1}\alpha}{2E_s} & \alpha\alpha^T \prec
\frac{E_s}{N_0}\Gamma\\
\infty & \mbox{elsewhere}\end{array}\right.
\end{eqnarray}
where the notation $A \prec B$, for two $k\times k$ matrices $A$ and $B$, means
that $B-A$ is a non-negative definite matrix.
The condition $\alpha\alpha^T \prec \frac{E_s}{N_0}\Gamma$ is equivalent to
$\alpha^T\Gamma^{-1}\alpha < E_s/N_0$. Here, the estimation error of the ML
estimator of $\theta$ is a zero mean Gaussian vector with covariance matrix
given by
$\frac{N_0}{2E_s}\Gamma^{-1}$, and so the exponential
moment performance of this estimator is given by
\begin{equation}
\Lambda_{\mbox{\tiny NB}}(\ct_{\mbox{\tiny ML}},\theta,\alpha)=
-\frac{1}{2}\ln\bigg|I-\frac{N_0}{E_s}\alpha\alpha^T\Gamma^{-1}\bigg|
=-\frac{1}{2}\ln\left(1-\frac{N_0}{E_s}
\alpha^T\Gamma^{-1}\alpha\right)
\end{equation}
which is larger than the lower bound but is asymptotically the same when
$\frac{N_0}{E_s}\alpha^T\Gamma^{-1}\alpha \ll 1$. Here $\alpha_{\mbox{\tiny
c}}$ is extended from a single point, in the scalar parameter case,
to the contour of a $k$--dimensional ellipsoid
defined by $\alpha_{\mbox{\tiny c}}^T\Gamma^{-1}\alpha_{\mbox{\tiny
c}}=E_s/N_0$, and so, the ML estimator has an optimal tail (in all directions) in
this sense.

\subsection{The Non--linear Signal Model}

Consider next the model (\ref{signalinoise}), where the energy
$E_x=\int_0^Tx^2(t,\theta)\mbox{d}t$ is independent of $\theta$.
We will also denote
\begin{equation}
\rho(\theta,\ttt)=\frac{\int_0^Tx(t,\theta)x(t,\ttt)\mbox{d}t}{E}.
\end{equation}
In this case, the lower bound becomes
\begin{equation}
\Lambda_{\mbox{\tiny NB}}(\ct_{\mbox{\tiny ML}},\theta,\alpha)\ge
\sup_{\ttt}\left[\alpha L_{\mbox{\tiny
NB}}(\ttt)+\alpha(\theta-\ttt)^2-\frac{2E[1-\rho(\theta,\ttt)]}{N_0}\right].
\end{equation}
If $L_{\mbox{\tiny NB}}(\ttt)$ is independent of $\ttt$ (which is nearly
the case with the Cram\'er--Rao lower bound 
in delay, phase and frequency estimation), and if the range $\calA$ of $\theta$
is unlimited, then $\ttt$ can be chosen such that the second term is
arbitrarily large (whereas the third term is bounded since
$|\rho(\theta,\ttt)|\le 1$), and so, the lower bound is infinite for every
$\alpha > 0$, namely, $\alpha_{\mbox{\tiny c}}=0$. In other words, in this
case, no unbiased estimator (if at all existent) has a Gaussian tail however
slow. If the range of $\theta$ is
limited, then, of course, the supermum is finite and it depends on the
exact form of $\rho(\cdot,\cdot)$. The same conclusion applies whenever the energy
of $x(t,\theta)$ grows slower than quadratically with $\theta$.
The conclusion that $\alpha_{\mbox{\tiny c}}=0$ is not surprising in view of
the fact that for the same signal model, we obtained $\alpha_{\mbox{\tiny
c}}=\frac{1}{2\sigma^2}$ in the Gaussian prior case. In the Bayesian case,
$\alpha_{\mbox{\tiny c}}$ was positive due to the prior only, and now the
prior does not exist anyway, so that $\alpha_{\mbox{\tiny c}}$ vanishes.

\end{document}